\documentclass[12pt,letterpaper]{JHEP3}
\usepackage{epsfig}
\usepackage{amssymb}
\usepackage{amsfonts}
\usepackage{amsmath}
\usepackage{amsbsy}
\usepackage{graphicx} 
\usepackage{amsmath}
\usepackage{amssymb}
\usepackage{subfigure}

\newcommand{\be}{\begin{equation}}

\newcommand{\ee}{\end{equation}}

\title{Making predictions in the multiverse}

\author{Ben Freivogel\\
Center for Theoretical Physics and Laboratory for Nuclear Science\\
 Massachusetts Institute of Technology, Cambridge, MA 02139, U.S.A.
}

\abstract{I describe reasons to think we are living in an eternally
  inflating multiverse where the observable ``constants'' of nature
  vary from place to place. The major obstacle to making predictions
  in this context is that we must regulate the infinities of eternal
  inflation. I review a number of proposed regulators, or measures.
  Recent work has ruled out a number of measures by showing that they
  conflict with observation, and focused attention on a few
  proposals. Further, several different measures have been
  shown to be equivalent. I describe some of the many nontrivial tests
  these measures will face as we learn more from theory,
  experiment, and observation.  }

\begin{document}

\section{Introduction}

Weinberg's successful prediction of the cosmological constant \cite{weinberg} suggests that we are living in a very
large universe where the ``constants" of nature vary from place to
place. In the years since Weinberg's prediction, advances in string
theory have lent theoretical support to this idea. While string theory
is a unique 11-dimensional theory with no free parameters, it seems to contain
a huge landscape of solutions with 4 large dimensions and 7 tiny
dimensions \cite{landscape}.  At the energy scales we can access,
these different solutions look like distinct 4-dimensional theories
with different values for the physical constants, such as the
cosmological constant $\Lambda$ and the Higgs mass. 

Furthermore, eternal inflation occurs naturally in the string
landscape. Starting from finite initial conditions, eternal inflation
produces an arbitrarily large spacetime volume in the inflating false
vacuum. Inflation ends locally, producing ``pocket universes" where
the fields settle into one of the vacua \cite{eternalinf}. Globally, inflation never
ends, and all of the vacua of string theory are produced as pocket
universes within a vast eternally inflating cosmology often called the $multiverse$. 

If the fundamental theory produces such a large universe where the
low-energy laws of physics vary from place to place, how are we to
make predictions? First, it is natural to use the anthropic
principle. In this context the anthropic principle is simply a
controversial name for a mundane selection effect: when we are
predicting the results of an observation, we can focus on the parts of
the multiverse where observations can occur. More precisely, if we had
complete knowledge of the multiverse, the relative probability of two
different observations is given by $counting$ the number of
observations of each type, \be
\label{eq1}
{p_A \over p_B} = {N_A \over N_B}~.
\ee
For example, to predict the value of $\Lambda$ we count the number of observations of different values in the multiverse.

In this setting, the constants we observe may be fine-tuned from the
conventional point of view. For example, $\Lambda$ is 123 orders of
magnitude smaller than its natural value. However, we expect our
observations will not appear fine-tuned among regions where observations occur.
Weinberg's prediction that $\Lambda$ would be detected soon was based on this assumption.   

Because computing the cosmological constant requires a theory of
quantum gravity, one could hope that the observed value of $\Lambda$
will turn out to be conventionally natural, without appealing to
anthropic selection in the multiverse. However, there
are a number of other parameters such as the charge of the electron
that appear fine-tuned for life, providing additional evidence for the multiverse.

It is disappointing that the fundamental theory gives only a
probabilistic prediction for quantities like $\Lambda$, because we
will never be able to measure the value of $\Lambda$ in other parts of
the multiverse. But our disappointment does not mean that the theory
is wrong. In fact, we already accept that a fundamental theory may
give only a probabilistic prediction for a quantity that we can only
measure once: for example, the our best theory of the early universe gives
only a statistical prediction for the $\ell = 3$ modes in the CMB. But
no one questions the moral integrity of the theory of slow roll
inflation as a result.  What is needed for
the theory of the multiverse to take its place as a well-accepted
physical theory is simply better theoretical control and
more opportunities to compare with experiment.

The major obstacle of principle to implementing the program of making
predictions by counting observations in the multiverse is the
existence of divergences. Eternal inflation produces not just a very
large universe, but an infinite universe containing an infinite number
of pocket universes, each of which is itself infinite. Therefore both the numerator and the denominator
of (\ref{eq1}) are infinite. We can define the ratio by regulating the
infinite volume, but it turns out that the result is highly
regulator-dependent.

There are two possible conclusions: either the proposal (\ref{eq1}) is
fundamentally ill-defined, or quantum gravity gives a way of defining
it.
It would seem that since string theory is a consistent theory of
quantum gravity, it should be able to answer if and how equation (\ref{eq1})
is defined. Unfortunately, this question is not yet tractable. 
We have an exact nonperturbative description of spacetimes with $\Lambda < 0$ that are
asymptotically Anti-de~Sitter (AdS) in terms of a dual conformal field
theory, the famous AdS/CFT correspondence. Similarly, Matrix theory is a nonperturbative description of
spacetimes with $\Lambda = 0$ and Minkowski asymptotics. But we do not
have the corresponding description of spacetimes with $\Lambda > 0$
and eternally inflating asymptotics; even more generally, we do not
have a  rigorous description of
$any$ cosmology. One reason for this difficulty is that the
asymptotic behavior of eternal inflation- more and more pocket universes in
the future- is much more complicated than the asymptotic behavior of
AdS spacetimes, which do not fluctuate near the boundary.

Because we lack the tools to address eternal inflation in a completely
rigorous way within string theory, our understanding necessarily
relies on approximations. It would be extremely interesting to develop
the necessary tools to conclusively establish the existence or absence
of the multiverse. 
An alternative is to take a more phenomenological approach, trying to
understand some predictions of the theory before it is completely
worked out.  I will mention some hints that quantum gravity
$does$ regulate equation (\ref{eq1}) in section
\ref{predictions}. This is fortunate because I am not aware of any
other proposal for how to make predictions
in the context of the landscape.

The first step in making any prediction in eternal inflation is to
regulate the infinities to make (\ref{eq1}) well-defined. A procedure
for regulating the infinities is called a $measure$.  I will
focus on {\it geometric cutoffs}: measures that supplement the
semiclassical treatment of eternal inflation with a 
prescription for cutting off the infinite spacetime volume.  In section \ref{predictions} I will describe a number of simple measure proposals and their
properties. Several of these proposals preserve a property that makes
eternal inflation particularly attractive for making predictions: the
late-time behavior is independent of the initial conditions, so all we
need to know about the initial conditions is that they allow
eternal inflation to occur. The late-time attractor behavior, however,
does depend on the choice of cutoff.

 In section \ref{deriving}, I will describe some steps towards deriving the measure from quantum gravity.
Fortunately, as I will discuss in section \ref{equivalences},
we have discovered surprising equivalences between proposals that sound
very different. Furthermore, some of the reasonable proposals conflict
strongly with observation and can be ruled out. This finally leaves us
with only two or three distinct proposals.

Then I describe, in section \ref{tests}, the many future tests that proposals will have to
pass, emphasizing that all extant proposals could easily be ruled out in the
near future.
Finally I conclude with a brief summary of the status of measure proposals.

As a prelude, in section 2 I describe a simple set of assumptions
leading to the conclusion that eternal inflation occurs in our past.

This is a personal review of the state of the field, reflecting my own
prejudices and ignorance. In particular, I have not made an effort to
cite every relevant paper on the subject. Where I have included citations I have tried to refer to useful references rather than original work. This may irritate
my friends who work on eternal inflation, but it will hopefully lead
to a more readable article. 

\section{Is eternal inflation in our past?}

It sounds contradictory to ask whether eternal inflation is in our past, since
eternal processes never end. But in the theory of eternal inflation,
some regions of spacetime do stop inflating. The question is whether a
long period of eternal inflation occurred before our pocket universe
formed.

In this section I will spell out a set of assumptions that leads to
the conclusion that eternal inflation is in our past.
  While many readers will find this a boring
exercise, some physicists believe the conclusion is obviously wrong. Therefore, I
think it is worth stating the assumptions clearly. Those who are
already convinced that eternal inflation is in our past can skip this section.

\paragraph{Assumption 1: The potential allows eternal inflation to occur.}
For simplicity, we focus here on eternal inflation that occurs in a metastable false vacuum. In order for the landscape to allow eternal inflation to occur, it must contain at least one false vacuum whose decay rate is slower than its Hubble expansion rate,
\be
\label{eibound}
\Gamma \lesssim H^4~.
\ee
Because the decay of a metastable vacuum is a nonperturbative process,
$\Gamma$ is naturally exponentially small. Further, string theory
seems to contain a very large number of metastable false vacua. It would take a vast conspiracy to avoid having at least one vacuum that satisfies the bound above.

\paragraph{Assumption 2: Initial conditions.}
Suppose the theory contains a false vacuum whose decay rate is slow enough to satisfy (\ref{eibound}). For eternal inflation to get started, we need to begin with several Hubble volumes that are (a) in the false vacuum and (b) are dominated by vacuum energy. The required initial conditions are generic in the technical sense: arbitrary small perturbations of the initial conditions will still allow eternal inflation to occur. Therefore the initial conditions that allow for eternal inflation form an open set in the set of all initial conditions.

Of course in a sense the initial conditions allowing for eternal inflation are very special.  We do not know the correct theory of inital conditions, so it is hard to say how special they are. What I will assume is that
 the theory of initial conditions gives a nonzero probability to begin in the open set of initial conditions that allows for eternal inflation to begin.

I also assume that the initial conditions are spatially finite. If the initial conditions are defined on an infinite spatial slice, we already have a problem of infinities before even considering the dynamics of eternal inflation.

\paragraph{Assumption 3: Typicality.}
In determining where in the multiverse we are living, we make the
assumption of typicality: we are equally likely to be anywhere
consistent with our data. This is called the ``principle of
indifference."

With our assumptions, there is a finite probability
for eternal inflation, which results in an infinite number of observations, so we can ignore any finite
number of observations.\footnote{This conclusion relies on an
  assumption about how to implement the typicality assumption when there is a probability distribution over how many observations occur \cite{page}. I
advocate first constructing the ensemble of probabilities and then
using typicality within that wider ensemble. Page calls this choice
{\it observational averaging}. As a simple example,
suppose our theory is that God flips a fair coin, and if it is heads
he makes one earth, while if it is tails he makes two earths that are
far apart. Suppose we are about to do some observations that will determine whether there is another earth out there. I conclude the probability of observing another earth is 2/3.  This turns out to be a
controversial conclusion among philosophers; it is one version of the
``sleeping beauty paradox.''}  Then to make predictions we can focus on the eternally
inflating branch of the wave function. Within this branch, again we
can ignore the finite number of observations that occur at early times. Thus with these assumptions a
long period of eternal inflation is in our past.

\section{Predictions in Eternal Inflation}
\label{predictions}
Having argued that we are living during the late time era of
eternal inflation, what are the predictions? If inflation were not
quite eternal, but just led to an extremely large spacetime, the
natural way to make predictions would be to count the number of events of
different types, as in equation (\ref{eq1}). However, this prescription becomes ambiguous if
inflation is truly eternal.

One possibility at this point is to conclude that the ratios we want
to compute are just not gauge invariant, and we are thinking about the
problem wrong. From the point of view of semiclassical gravity this is
the obvious conclusion because no principle within the theory gives a
preferrred way of defining (\ref{eq1}).

However, there are some hints that in quantum gravity the ratio $N_A /
N_B$ may be
well-defined. The infinities causing the ratio to be ill-defined come
from counting events in causally disconnected regions of spacetime. We
have learned from studying black holes that attempting to use
semiclassical quantum gravity in causally disconnected regions of
spacetime can lead to confusion. In the case of black holes,
semiclassical analysis led to the conclusion that black holes destroy
information. Even though the analysis seemed to be in a regime of low
curvatures where semiclassical gravity is a good approximation, we now
know that in a full theory of quantum gravity evolution is unitary.

The lesson I and many others take away from black hole physics is that
semiclassical gravity can be trusted only within a single {\it causal
  diamond}. For a given worldline, the causal diamond is the region of
spacetime that can send signals to and receive signals from the
wordline; it is the largest region that can be probed in principle by
a single observer.  

In de Sitter space,
the exponential expansion causes spatially separated points to fall
out of causal contact with each other.
The infinities of eternal inflation arise from these spacetime regions that
are out of causal contact with each other, so the analogy with black holes suggests that the infinities are figments of
our semiclassical imaginations\footnote{An exception is the causal patch of a
  worldline that enters a Minkowkski vacuum, which can contain an
  infinite number of observations. I will return to this
  example later.}. 
Given this encouragement and a dearth of other proposals, we
will pursue the idea that eternal inflation is the right machine for
making predictions from the string theory landscape. 

\subsection{Local Measures}
\label{local}
\paragraph{The causal diamond measure.} 
The causal diamond cutoff of
Bousso \cite{cd} is motivated by the lessons of black hole
physics. This cutoff keeps only those events occuring within a single causal
diamond.

One still must specify the initial conditions for the diamond. The simplest option is to say that the specification of
initial conditions is a separate problem from the
measure problem.
Another possibility is to define a rule for going from the global
picture of the eternally inflating spacetime to an ensemble of causal
diamonds.

The most basic question is whether the causal diamond cutoff succeeds
in regulating the infinities. The answer is yes, as long as the proper
time along the central worldline is finite; then the volume of the
causal diamond will be finite. However, for any reasonable choice of
initial conditions there is a nonzero probability for a worldline to
tunnel to a supersymmetric $\Lambda = 0$ bubble. Once inside the
$\Lambda = 0 $ bubble, it is believed that there is a nonzero
probability for the worldline to attain infinite proper time \cite{insidestory}.
These infinite worldlines lead to divergences, as Bousso already realized in his
original paper, so the measure is defined to count only worldlines of
finite length.

The causal diamond cutoff regulates the infinities of eternal
inflation so strictly that the late-time attractor behavior disappears as
well. The number of events of different types computed according to
the causal diamond cutoff depends on the choice of initial
conditions. So this cutoff is NOT a prescription for understanding the
attractor behavior of eternal inflation. Instead, it tells us that the
attractor behavior is an artifact of the same global picture that gave
us infinity in the first place.

There are two simple variations on the causal diamond cutoff that are
not quite as well motivated from black hole physics:
\begin{itemize}
\item{ {\bf The apparent horizon measure.}  This cutoff includes those events within the apparent horizon of the central geodesic \cite{general}, rather than keeping the entire causal diamond. This measure is very similar to the causal diamond measure and has not received as much attention, so I will not discuss it further here.}

\item{ {\bf The fat geodesic measure.} Finally, instead of keeping all events within the causal diamond, one can only keep those events within a fixed physical volume centered on the geodesic \cite{ussf}. This proposal was motivated by an equivalence to a global cutoff, and we will discuss it more in the next section.}
\end{itemize}

\paragraph{The census taker cutoff.}
An alternative to throwing out the causal diamonds that become
infinitely large is to focus on them. The ``census taker cutoff'' is
the most famous unpublished cutoff prescription \cite{ct}.
Consider a worldline that ends in a hat, and therefore has infinite
length. Counting everything within the causal diamond of this
worldline is less infinite than the entire global multiverse, but it
is still infinite. However, suppose we just count all events that can
send a signal to the central worldline before proper time $\tau$. This
is a finite set. We can now take the limit $\tau \to \infty$.

It is plausible that the probabilities defined in this way are
independent of initial conditions. This prescription has the advantage
of keeping the attractor behavior of eternal inflation while
restricting to a single causally connected region. As far as anyone
knows, the census taker cutoff is not ruled out by observation.

However, there is one major aesthetic problem that accounts for the
fact that it has not been published. Imagine that we are trying to
send a signal to the census taker. We need to arrange to live in a
region of spacetime near the $\Lambda = 0$ bubble. We want to
send a signal through the domain wall that separates us from the
$\Lambda = 0$ bubble.

It turns out that our ability to be counted in the census depends
crucially on the behavior of the domain wall between our bubble and
the $\Lambda=0$ bubble.  Because the $\Lambda = 0$ bubble has smaller
vacuum energy, we know that a free falling observer inside the
$\Lambda=0$ bubble will see the domain wall accelerate away from
him. However, it can still accelerate towards us or away from us; it
is a counterintuitive feature of general relativity that the domain
wall can appear to accelerate away as seen from both sides. If the
domain wall tension is low enough, it will accelerate towards
us. In this case it is not difficult to be counted, because the domain
wall could well come into our future lightcone.

However, if the tension is too big, then the domain wall will
accelerate away from us, and the region of our vacuum that can send
signals to the census taker will have only a microscopic
thickness. For our value of the cosmological constant, the critical
tension that divides these two behaviors is 
\be
T \sim {  \sqrt \Lambda  \over G_N} \sim (\rm GeV)^3
\ee
It seems absurd that whether we are counted in the census depends on
such details as the tension of a particular domain wall. To put it
another way, the census taker cutoff would predict that the domain
wall between our vacuum and a supersymmetric $\Lambda = 0$ vacuum is
very likely to have a small tension. This dependence on arcane details
makes the census taker cutoff aesthetically unattractive.

The census taker description of eternal inflation \cite{census} still may be a
valuable one, but probably not in the simple sense of only counting
events that take place within the census taker's backward lightcone.

\subsection{Global Measures}
\label{global}
Historically, the local measures were only developed long after the
global time cutoffs. The idea of a global time cutoff is to pick some
preferred global time variable in the multiverse. We first count only
events that happen before some cutoff time $t_0$, then take the limit
$t_0 \to \infty$. Assuming the initial conditions are finite, there is
only a finite spacetime volume before a finite time $t_0$, so this
procedure succeeeds in regulating the infinities; in all known cases
the limit $t \to \infty$ is well-behaved.  Several of these time
variables were introduced by Linde and collaborators in 1993
\cite{linde93}; in 1995 Vilenkin proposed using a global time variable
as a cutoff for computing probabilities \cite{vil95}.

\paragraph{The proper time cutoff.} Perhaps the simplest definition is
to use the proper time as a cutoff \cite{linde93}.  Begin with a finite spacelike initial
surface $\Sigma_0$ that we will define to be $\tau = 0$, and erect a congruence of timelike geodesics orthogonal to
that surface. The time at some other point is given by the proper time
from the initial surface, measured along the geodesic in the
congruence that connects the given point to the initial surface. We
first count all events before some proper time $\tau_0$, then take the
limit $\tau_0 \to \infty$. The resulting probabilities will be
independent of the initial conditions.

There are various possible technical issues with this definition; for
example, we have to decide how to define the time if there are two geodesics connecting the given point to
the initial surface, as will happen to the future of caustics. These
issues are not worth worrying about because this proposal has more
severe problems: namely, it suffers from the ``youngness problem''
\cite{center, youngness}.
This cutoff predicts that we are incredibly unlikely to live at such a late
time, 13.7 billion years after our local big bang. The probability to live only, say, 13 billion years after the big bang is larger by the enormous factor $\exp \left(10^{60} \right)$ \cite{youngness}. Therefore, the proper time cutoff conflicts with observation.

Linde and collaborators have attempted to modify the proper time
cutoff to resolve this conflict with observation \cite{fixes}. I personally
have not been able to understand from their work a relatively simple, well-defined
modification that brings the cutoff into agreement with observation.

\paragraph{The scale factor time cutoff.}
The scale factor cutoff is defined similarly to the proper time
cutoff, by considering the same geodesic congruence orthogonal to an initial
surface $\Sigma_0$. Now, however, the time is measured by the expansion
rather than by the proper time. Along a geodesic, the scale factor
time is defined by
\be
\label{sf}
dt = H d\tau
\ee
where $\tau$ is the proper time and $H$ is the local expansion of the
geodesic congruence. Roughly, this means that time advances by one
e-folding everywhere.
This cutoff first appeared (as far as I know) in the work of Linde and collaborators in 1993 \cite{linde93}, and was first defined carefully by De~Simone, Guth, Salem, and Vilenkin in 2008 \cite{scalefactor}.

Again, the definition (\ref{sf}) brings up various technical questions. In this
case I will mention them because they are the worst aspect of this cutoff procedure. The main issue is that the above definition becomes ambiguous in the future of caustics. In the future of a caustic, defining the time by the above equation is not unique because there is more than one geodesic leading to the point under consideration. \cite{scalefactor} made a choice for what to do, but a result is that in order to compute the current scale factor time we need to understand the intricacies of geodesic motion around our galaxy \cite{ussf}. 
The proposal does not conflict with observation as far as we know, but it just seems wrong that the predictions should depend on the gory details of the motion of geodesics in our galaxy.
\cite{guthbb} suggested one way of modifying the proposal to avoid
this issue.

\begin{figure}[tbp]
\label{lightconefig}
\centering
 \includegraphics[width=5 in]{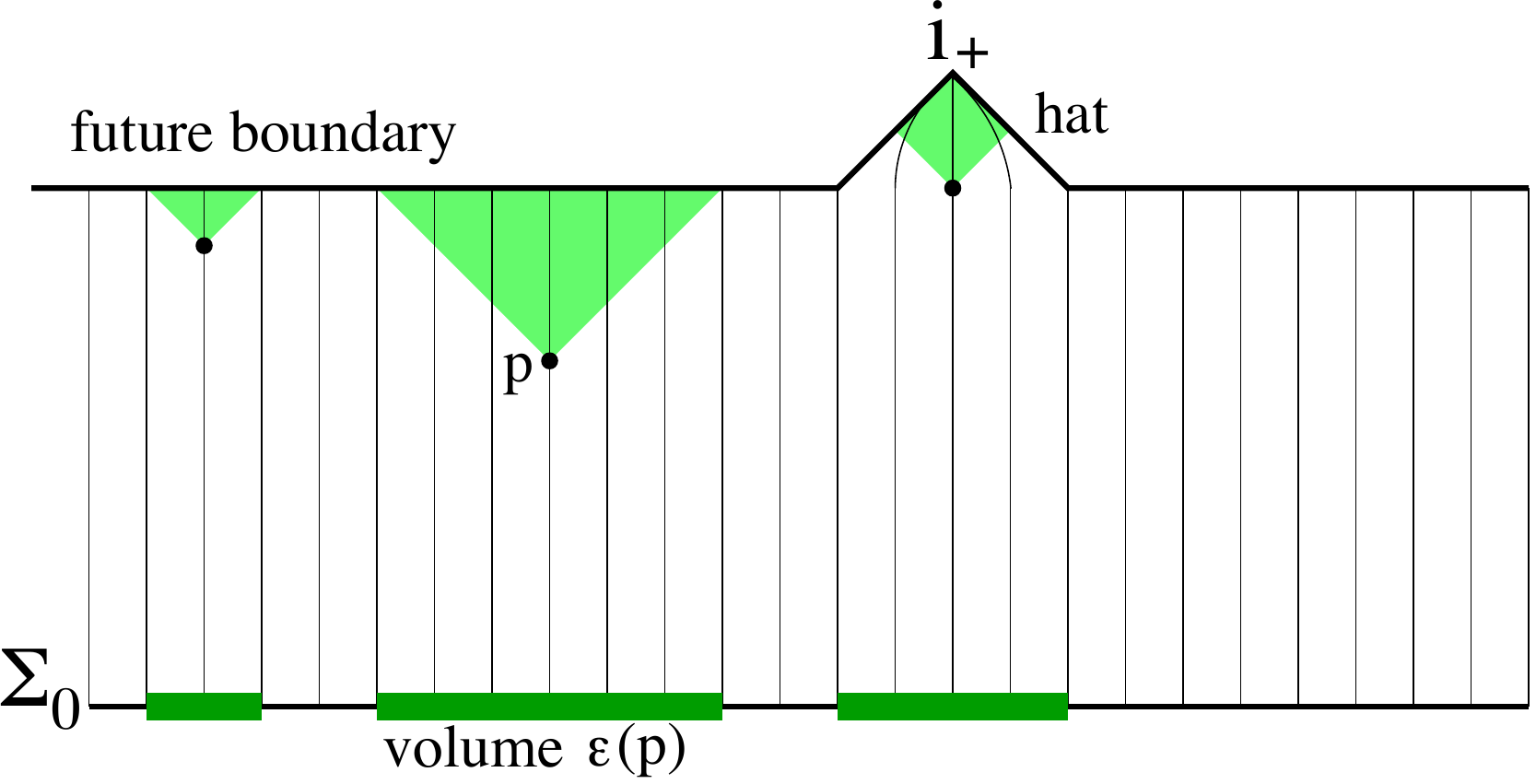}
\caption{To compute the lightcone time of an event, construct its
  future lightcone and project back to the initial surface $\Sigma_0$
  along the geodesic congruence. The resulting volume on $\Sigma_0$
  gives the lightcone time of the event. Figure courtesy of Raphael Bousso.}
\end{figure}

\paragraph{The lightcone time cutoff.}
The lightcone time cutoff makes use of the same geodesic congruence described above. To find the time at some spacetime point, construct its future lightcone. The future lightcone will capture some of the geodesics, as shown in figure \ref{lightconefig}.
 Following these captured geodesics back to the initial surface, find
 the volume $\epsilon$ of the captured geodesics on the initial surface. The lightcone time is related to the volume by
\be
t \equiv - {1 \over 3} \log \epsilon~.
\ee

This time variable was first defined by Garriga {\it et al.}
\cite{villc} in
the more restricted context of counting the number of bubbles of each
type. In that context it is extremely natural to consider the
geodesics in the future lightcone because bubbles expand out from the nucleation point at the
speed of light. It was first proposed as a cutoff- that is, as a rule for
counting any type of event- by Bousso \cite{lightcone}. 

There is a sense in which the lightcone time is better defined than
the scale factor time: no ambiguity arises in computing the volume
when the
geodesic congruence has caustics. Similarly, the lightcone time of an
event depends only weakly on details of geodesic motion since most of
the geodesics in the future lightcone never enter galaxies. (See
however \cite{albrecht} for possibly significant effects of structure formation.)

On the other hand, Vilenkin has complained that lightcone time suffers
from a problem he calls ``shadows of the future'': the lightcone time
of an event depends on what happens in the future of that event,
because future events can affect the size of the lightcone. 

\section{Towards deriving a cutoff}
\label{deriving}

We would prefer to be able to derive the cutoff directly from
theory. An obstacle to this is that the rigorous description of
eternal inflation in string theory is not known. But we do have some
idea for what a rigorous description might look like, and even without
fully understanding the theory we can make some progress towards
deriving a cutoff prescription.

We do not believe that completely stable de Sitter space is possible
in string theory. If it were, it would be natural to assume that there
is a dual, nongravitational theory living on the conformal boundary of
the spacetime \cite{dscft}. Looking
at holographic entropy bounds \cite{sbound} informs us that future
infinity is a holographic screen for de Sitter spacetime, and thus the
natural place for the dual theory.

The fact that the dS/CFT correspondence has not been extremely
successful is probably due to the problem that completely stable de
Sitter space does not seem to exist in string theory. However, a dual
description on future infinity still seems natural for the part of the
spacetime that is still in the false vacuum. 

\begin{figure}[tbp]
\subfigure[$\Lambda>0$]{
   \includegraphics[width=3in]{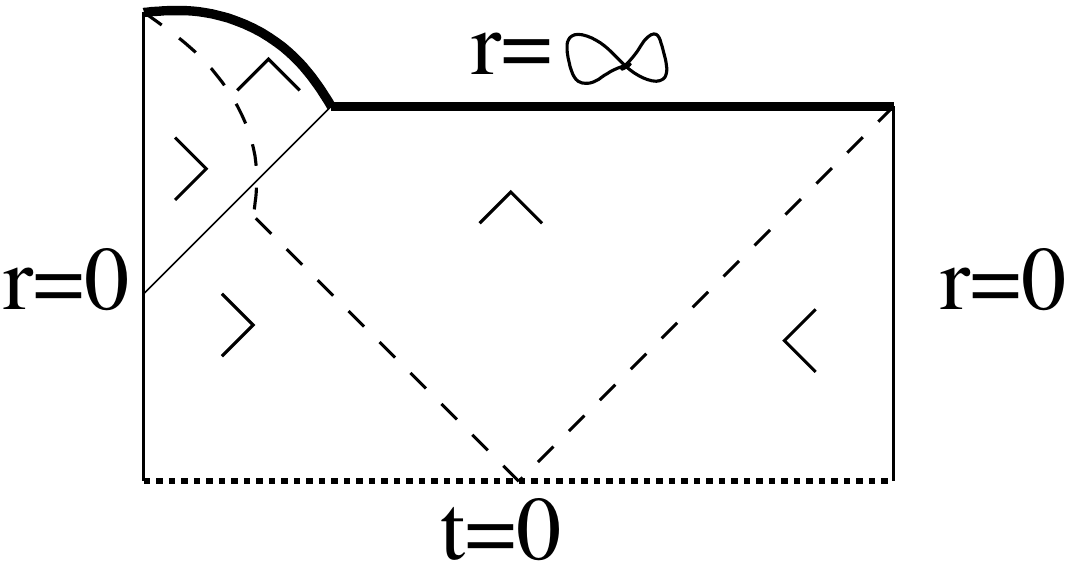}
   }
    \subfigure[$\Lambda<0$]{
    \includegraphics[width=3 in]{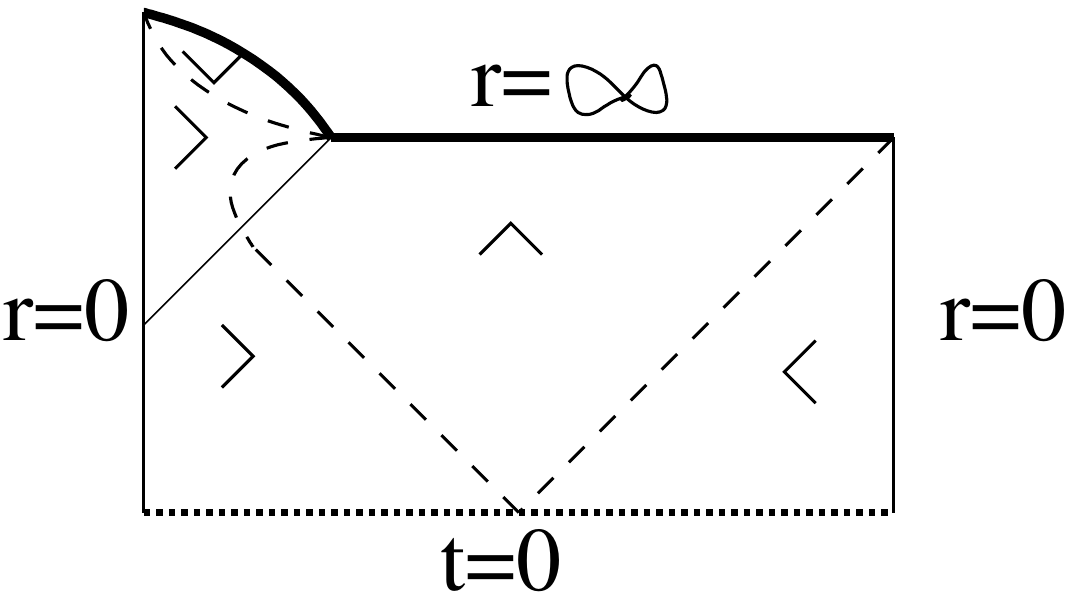}
    }
   \caption{Conformal diagrams for bubbles with positive and negative
     cosmological constant. The Bousso wedges indicate the null
     directions in which the sizes of spheres are increasing. The tips
     of the wedges point in the direction of increase. Entropy bounds
     dictate that if we think of the wedges as arrows, they point
     towards the natural holographic screens \cite{sbound}. One can
     see that future infinity is a natural holographic screen for the
     $\Lambda>0$ bubble (left), but not for the $\Lambda<0$ bubble.}
   \label{confdiags} 
\end{figure}

 The main theoretical problem has to do with vacua with
$\Lambda \leq 0$. For bubbles with $\Lambda \leq 0$, it is not natural
to expect that the bulk physics is captured by a dual description that
lives on future infinity. This can be seen from the conformal diagrams
in figure \ref{confdiags}.
It seems that these bubbles make holes in the theory on
future infinity. This is of course not something that can happen in a
conventional field theory; it suggests that the metric of the boundary
must fluctuate. There were already hints of this perturbatively: de
Sitter space inevitably has gravitational fluctuations that freeze out
on scales larger than the horizon, so the natural state in the bulk
corresponds to a sum over boundary geometries.

Given these vague outlines of a dual theory, we can try to derive a
cutoff. The basic idea is the UV/IR correspondence familiar from
AdS/CFT: in this case a short-distance (UV) cutoff in the boundary
theory corresponds to a late-time (IR) cutoff in the bulk. This
late-time cutoff renders the spacetime volume finite and therefore
regulates the infinities. The first proposal for a measure derived in this way was made by Garriga and Vilenkin \cite{gv}.

\subsection{The new lightcone time cutoff}
The first step is to define a UV cutoff in the boundary theory. 
In order to how to regulate the boundary theory in the UV, we first need to know the metric on the boundary. In the physical spacetime, the
metric diverges at future infinity. To define the metric on future
infinity, we do a conformal transformation of the bulk
spacetime. However, there is no unique way to choose the ``right''
conformal transformation. The result is that, for a given physical
spacetime, the boundary metric is only defined up to conformal transformations,
\be
\tilde g_{ab} = e^{2 \phi} g_{ab}
\ee
where $\phi$ is an arbitrary function. This is a problem, because in
order to put in a uniform UV cutoff in the boundary theory, we need to
know the metric. There is no fiducial choice, and different natural
choices will correspond to physically different cutoff prescriptions.

At this point it would be nice to know more about the boundary
theory. But in our ignorant state, we can hypothesize a simple,
natural way to fix the conformal factor of the boundary metric: we
demand \cite{yamabe}
\be
^{(3)} R = {\rm const.} \ \ \ \ \ \ ^{(3)}V = 1~.
\ee
Some nontrivial mathematics guarantees that this choice can be made
and generically completely fixes the ambiguity.

This choice of metric is less arbitrary than it seems at first. First
of all, as long as we demand the scalar curvature remain finite, we
would get the same physical predictions. Second, Vilenkin has argued
recently \cite{vilenkinr} that the only sensible conditions for fixing the metric
are functions of the scalar curvature and its derivatives.

\paragraph{The bulk-boundary mapping.}
Having defined a UV cutoff in the boundary theory, we now need to know
how to relate this to an IR cutoff in the bulk. 
Even in the better-understood case of AdS/CFT, it is highly nontrivial
to work out how a given UV cutoff in the boundary theory translates
into the bulk. I will continue to focus on
bulk cutoffs that are completely geometric: a sharp cutoff surface
bounds the events that are counted.  Any simple cutoff in the boundary
theory will probably not correspond to such a simple bulk cutoff-
instead, the description of bulk events will grow fuzzy as they
approach the cutoff.

One simple recipe is
the following: given a bulk spacetime point, construct its future
lightcone. Find the volume of the region on the future boundary that
is contained inside the future lightcone. Keep only those bulk points
whose future lightcones have volumes bigger than some minimum
size. This can be stated by defining a time variable \be t \equiv - {1
  \over 3} \log V \ee where $V$ is the volume on future infinity that
is inside the future lightcone. The new lightcone time is clearly very
similar to the old lightcone time. The only difference is that in the
old lightcone time the volume is measured on the initial surface
$\Sigma_0$, while in the new lightcone time it is measured on future
infinity.

Having defined a time variable- the ``new lightcone time''- we can use
this as a regulator as before. We will return in the next section to further properties of this
measure. 

It is also possible to consider other ways of mapping a UV cutoff on
the boundary to an IR cutoff in the bulk. Instead of using the future
lightcone, Garriga and Vilenkin \cite{gv} consider trying to describe
a bulk process with minimum resolution $\lambda_{\rm min}$. The
boundary size is given by propogating the minimal length along
geodesics up to future infinity. Then we only count events whose
boundary size is bigger than the UV cutoff.  It is not completely
clear how to choose the scale $\lambda_{\rm min}$ in general. For
example, if we want to count observations of the cosmological
constant, what minimum resolution should we demand? On the other hand,
this proposal encodes a property one would intuively expect: It seems
reasonable that ants will cease to be resolved in the cutoff theory before
people due to their smaller size and mass.

More recently, Vilenkin has argued that the most natural bulk/boundary
mapping leads to a cutoff on bulk surfaces with constant comoving
apparent horizon \cite{vilenkinr}. 

We should not be surprised that different authors have come to
different conclusions about the best way to perform the bulk/boundary
mapping. Even in the much better understood case of AdS/CFT, the
bulk/boundary map is only simple on length scales large compared to
the curvature radius. Some progress has been made in describing
smaller objects in AdS (see, for example, \cite{berenstein}), but the
UV/IR relation becomes much more complicated. In particular, a simple
UV cutoff in the boundary theory probably does not correspond to a
sharp IR cutoff on the geometry in the bulk except on scales larger
than the curvature radius; see \cite{holorg} for recent progress in
addressing this question.

The key point is that how exactly the bulk spacetime is cut off in the
IR depends on the $details$ of the UV cutoff in the boundary
theory. Different UV cutoffs will give different prescriptions for
cutting off the bulk. If the boundary theory is a conventional field
theory, then the only quantities that make sense are those that are
independent of the details of the cutoff. So if the boundary theory
were a conventional field theory, we would have to conclude that the
bulk quantities we are computing do not make sense, because they
depend on the details of the cutoff. Fortunately, as we have
discussed, there is evidence that the boundary theory is not a
conventional field theory. It may be a theory that has a built-in UV
cutoff. 

In order to really fulfill our dream of deriving a cutoff and bring
our subject onto firm theoretical ground, we will have to make
progress in understanding the boundary theory, or whatever the correct
description of eternal inflation in quantum gravity is.

\section{Equivalences between measures; global-local duality}
\label{equivalences}

It is annoying to have so many reasonable-sounding measure
proposals. One encouraging fact is that many of these proposals turn
out to be equivalent to each other. First, as the terminology
suggests, the new lightcone time cutoff is equivalent to the old
lightcone time cutoff in the approximation that the bubbles are
homogeneous FRW universes \cite{yamabe}.

A more surprising correspondence has been discovered between the local
measures and the global measures \cite{lightcone, isheng, ussf},
called global-local duality. Recall that the local measures (except
the census taker, which we will not discuss further)
depend on initial conditions. The statement is that the global
measures are equivalent to local measures  $if$ the initial conditions
for the local measure is given by the attractor behavior of the global
measure. For the measures under consideration, this initial condition
is extremely simple: the geodesic should start in the most stable
vacuum with positive cosmological constant.

Given this choice of initial condition, the lightcone time cutoffs,
which are defined globally, are equivalent to the causal diamond
cutoff \cite{lightcone, isheng}. It is very encouraging that the new
lightcone time, which we motivated by the UV/IR correspondence, turns
out to be equivalent to counting events within a single causal
diamond.

Similarly, the scale factor cutoff is equivalent, in the approximation that the
geodesic congruence never stops expanding, to the ``fat geodesic''
cutoff \cite{ussf}.

So all of the measures discussed above reduce to only
two proposals that are still in agreement with observation: the
lightcone time cutoff and the scale factor cutoff. (We could count the
global and local versions of the
apparent horizon cutoff as a third possibility, but since these proposals are
rather new and not radically different I will ignore them for
brevity.) 

So it all boils down to this: consider a geodesic that begins in the
most stable de Sitter vacuum. Keep either all events within its causal
diamond (lightcone time/ causal diamond), or within a fixed physical volume orthogonal to the geodesic
(scale factor time/ fat geodesic). 


\section{Tests of measures and the landscape}
\label{tests}

There are many opportunities for measure proposals to conflict with
observation. As described above, the proper time measure is very
natural theoretically, but makes a completely wrong prediction about
the observed age of the universe. There are many additional tests, and
I will only briefly describe a few. The existing
measure proposals pass these tests, as far as we know given our
current knowledge of the landscape. However, as we learn more both
theoretically and experimentally, it could easily happen that all
the measures I have described here will be ruled out.

\subsection{Particle Physics}
One area where our knowledge of the landscape has so far limited our
ability to make predictions is particle physics. We would like to be
able to predict the supersymmetry breaking scale, among other
quantities. One could easily imagine a scenario where the landscape predicts
that the SUSY breaking scale is very high, near the Planck scale,
while the LHC reveals low-energy SUSY. This type of development has
the potential to rule out the entire multiverse framework for making
predictions, and the SUSY breaking scale is just one of many tests of
this type. Unfortunately, serious technical progress is needed before
we can extract these predictions from the landscape.

\subsection{Predicting the cosmological constant}
The most persuasive piece of evidence for the landscape is Weinberg's
prediction of the cosmological constant. However, there is also an
opportunity for measures to fail once we generalize Weinberg's
analysis and try to predict the cosmological constant allowing more
parameters to vary. We argued recently \cite{general} that for positive cosmological
constant both the scale factor and lightcone measures are very
successful in predicting the observed value of $\Lambda$, addressing
concerns that Weinberg's prediction is not robust when other
parameters are allowed to vary.

However, both measures are in danger of predicting that we are much
more likely to observe a negative cosmological constant than a
positive one, a conclusion already reached by Salem \cite{salem} and
by Bousso and Leichenauer \cite{bl} in specific cases.

\subsection{The Boltzmann Brain problem}
Despite the fanciful sounding name, the Boltzmann Brain problem is a
serious problem that has ruled out measures in the past \cite{bbs} and poses a
threat for currently popular measures as we learn more about the landscape.

The issue is that there are two ways observers can form in the
multiverse. The first is the traditional way: a pocket universe
forms, which then undergoes slow roll inflation, reheating, and
structure formation. This process produces a large
universe filled with matter. The second way structure can form is by a vacuum fluctuation
in de Sitter space. De Sitter space has a finite temperature, so
starting in empty de Sitter space, a fluctuation with mass $M$ occurs
with a rate given by
\be
\Gamma \sim H^{-1} \exp\left(- {M \over T}\right)
\ee
where the de Sitter temperature is $T =  H/(2 \pi)$. These
fluctuations violate the second law, and they produce a mass in a
universe that is otherwise completely empty.

As a concrete example of such a fluctuation, let us compute the expected time
to fluctuate the earth out of our de Sitter vacuum. Let us specify
that we want to fluctuate the earth in exactly its current state. The
time for such a crazy fluctuation is
\be
t \sim (10^{10} {\rm years}) \times  e^{\displaystyle 10^{92}}~.
\ee
This is an unimaginably long time; however, it is far shorter than the
recurrence time of our vacuum
\be
t_{\rm rec} \sim (10^{10} {\rm years}) \times  e^{\displaystyle 10^{123}}
\ee
Therefore, if our vacuum lives for of order the recurrence time, the
number of ``Boltzmann Earths'' that fluctuate out of the vacuum within
one causal patch is enormous, $N_{BE} \sim \exp(10^{123})$. This is
superexponentially more than the number of planets that form within one
causal patch by traditional structure formation; the number of
``ordinary earths'' is $N_{OE} \sim 10^{22}$.

If we focus attention just on our vacuum, and assume we are typical,
then we conclude that if the lifetime is of order the recurrence time,
we should be living on a Boltzmann Earth. But this conflicts with
observation, because observers on Boltzmann Earths are living in an
otherwise empty universe.

In the multiverse, observers form both in the ordinary way (``ordinary
observers'') and from fluctuations (``Boltzmann Brains''). In order to
agree with observation, it is important that the Boltzmann
Brains do not vastly outnumber the ordinary observers. For the
measures under consideration, the Boltzmann Brains will dominate if
$any$ vacuum in the landscape has a decay rate that is slower than its
rate for producing
Boltzmann Brains \cite{bbs, ussf, guthbb}. That is, to agree with
observation, every vacuum must satisfy
\be
\Gamma_{\rm decay} > \Gamma_{BB}
\ee
This is a highly nontrivial bound; for example, it
demands that our vacuum decay far faster than the recurrence time. As
far as we know, the landscape satisfies this nontrivial bound
\cite{lippert}, but we could find out otherwise any day.




\subsection{Alarming implications of geometric cutoffs: The end of
  time}

There is a sense in which all geometric cutoffs of eternal inflation
predict a novel type of catastrophe: we could run into the cutoff, and
time would end. I will describe the physics of the situation, but in
the end it is a matter of judgment whether one should conclude that
all geometric cutoffs are unsatisfactory. Predicting that time could end sounds
crazy, but it does not contradict observation if the probability of
encountering the end of time is small.

The main issue is this: in all cutoff prescriptions, a finite fraction
of the observers who are born before the cutoff run into the cutoff
before they die (see figure 3). This is true even for cutoffs that involve taking a
late time limit due to the exponential growth of the spacetime. An
analogy that is mathematically precise is a population that grows
exponentially until doomsday.  A finite fraction of everyone who has
ever lived is alive on doomsday. This fraction does not go to zero as
doomsday is taken later and later.

\begin{figure}[tbp]
\subfigure[]{
   \includegraphics[width=3in]{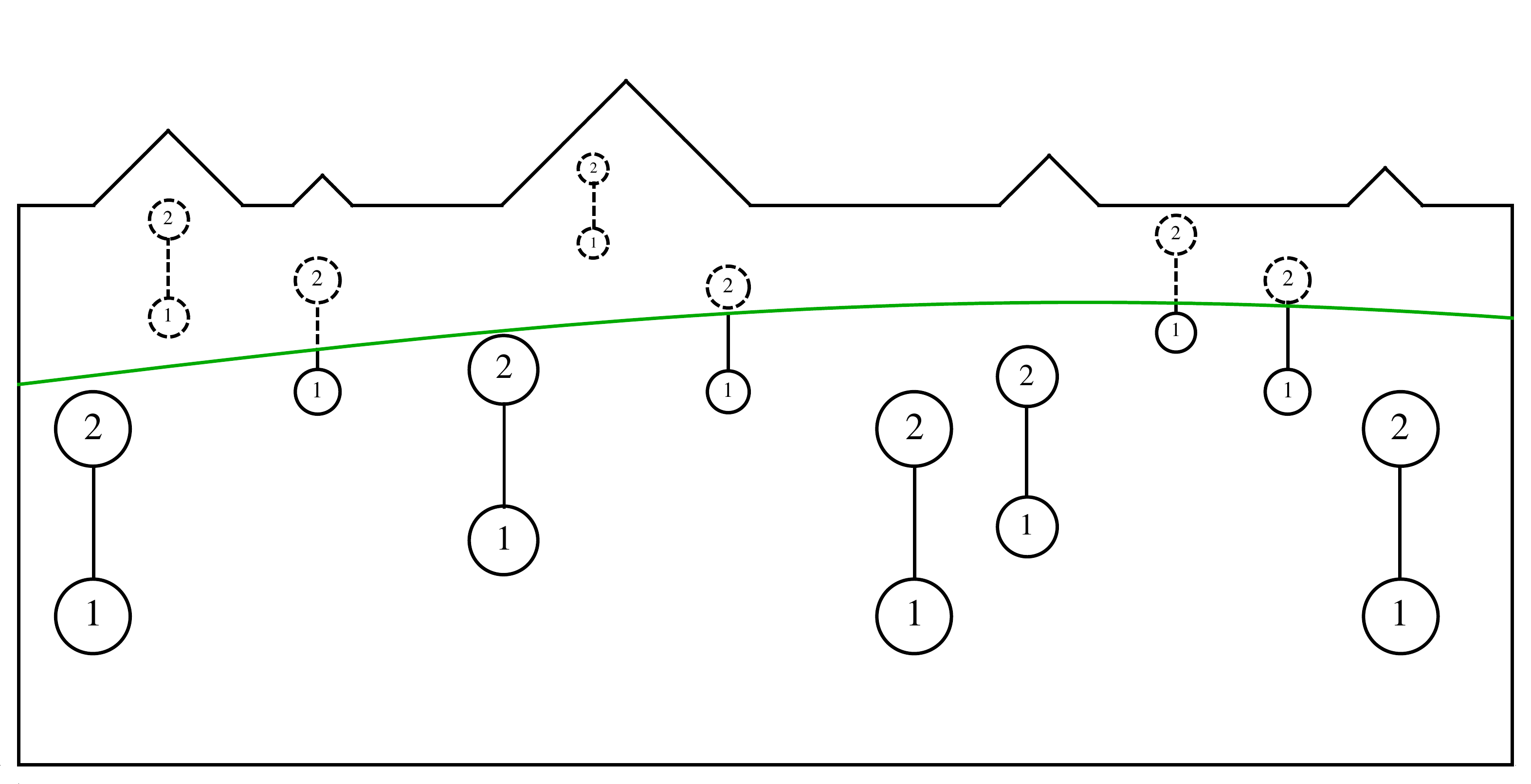}
   }
    \subfigure[]{
    \includegraphics[width=3 in]{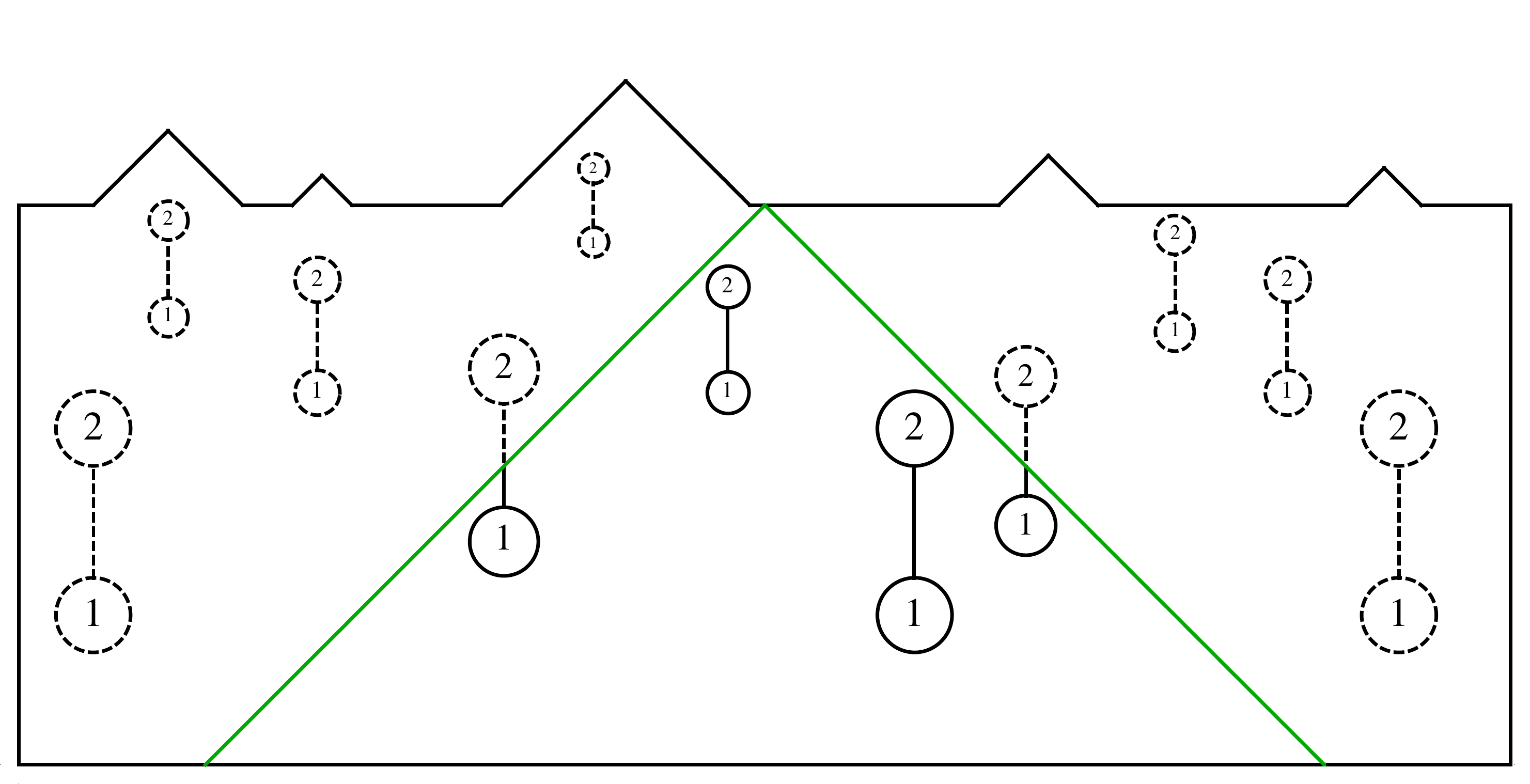}
    }
   \caption{In both global (left) and local (right) cutoffs, a finite
     fraction of observers born before the cutoff run into the cutoff
     before they die. The fraction of observers who are cut off does
     not go to zero as the cutoff is taken later and later.}
   \label{eotfig} 
\end{figure}

Because a finite fraction of the events happen close to the cutoff, we
are forced to give some physical interpretation to observers who run
into the cutoff. (If the population grew slower than exponentially,
then as the cutoff was taken later and later the fraction of observers
who run into the cutoff would go to zero, and we could forget about
them.)

One can try to think of the cutoff as just a mathematical device for
defining probabilities in an infinite set \cite{guthvanch}, and deny that the
cutoff is a physical entity we could run into. But it is not so easy
to escape the unpalatable consequences of the cutoff. The famous
Guth-Vanchurin  paradox  \cite{guthvanch, eot} illustrates this by showing that certain
probabilities computed in the multiverse conflict with common sense
expectations unless we take the end of time seriously as a catastrophe
that could happen to us.




\section{Summary: the status of measures}

Our efforts over the last several years have focused attention on two
simple measure proposals: the lightcone time cutoff and the scale
factor time cutoff. These proposals can be described in a very simple
way: follow a geodesic that begins in the most stable de Sitter vacuum
in the landscape. The lightcone time cutoff counts only those events
that are within the causal diamond centered on the geodesic, while the
scale factor cutoff counts only those events within a fixed physical
volume surrounding the geodesic.

Both proposals have passed a number of nontrivial
tests, but may be ruled out in the near future as we learn more about
the landscape.

There are two issues about these measures that concern me. The first
is the end of time issue described above. While the measures agree
with observation, predicting the end of time when there is no obvious
physical mechanism seems wrong. On the other hand, avoiding the end of
time conclusion seems to require a radical change in how we think
about the measure problem.

There is also a more concrete issue: the fact that these measures do
not seem to work well for negative cosmological constant. They have a
strong tendency  to predict we should observe $\Lambda < 0$, as I
described in the previous section. Because of theoretical
uncertainties, it is not yet clear that there is a strong conflict
with observation, but there are clear hints.

Developing a more rigorous understanding of eternal inflation in
string theory is crucial to putting this subject on firmer
footing. We are beginning to have the vague outlines of a dual
description for regions with $\Lambda > 0$, but we know very little
about the correct description of regions with $\Lambda < 0$. The
recent proposal of Maldacena \cite{malda} for a dual description of
crunches is very interesting, but it does not seem to describe well
realistic cosmologies with a period of slow roll inflation.

The future is very exciting. We can look forward to attacking two of the biggest theoretical
questions: the string theoretic description of cosmology and eternal
inflation, and how to extract predictions from string theory. At the
same time, our measures of the multiverse will continue to confront
experiment and observation.

\end{document}